\documentclass[twocolumn,showpacs,preprintnumbers,a4,prb]{revtex4}
\usepackage{graphicx}
\usepackage{dcolumn}
\usepackage{amsmath}
\usepackage{amssymb}
\usepackage{bm}

\begin{document}

\title{Influence of randomly distributed magnetic nanoparticles
on surface superconductivity in Nb films}
\author{D. Stamopoulos, M. Pissas, V. Karanasos, and D. Niarchos}
\affiliation{Institute of Materials Science, NCSR "Demokritos",
153-10, Aghia Paraskevi, Athens, Greece.}
\author{I. Panagiotopoulos}
\affiliation{Department of Materials Science and Engineering,
University of Ioannina, Ioannina 45110, Greece.}
\date{\today}

\begin{abstract}
We report on combined resistance and magnetic measurements in a
hybrid structure (HS) of randomly distributed anisotropic CoPt
magnetic nanoparticles (MN) embedded in a $160$ nm Nb thick film.
Our resistance measurements exhibited a sharp increase at the
magnetically determined bulk upper-critical fields $H_{\rm
c2}(T)$. Above these points the resistance curves are rounded,
attaining the normal state value at much higher fields identified
as the surface superconductivity fields $H_{\rm c3}(T)$. When
plotted in reduced temperature units, the characteristic field
lines $H_{\rm c3}(T)$ of the HS and of a pure Nb film, prepared at
exactly the same conditions, coincide for $H<10$ kOe, while for
fields $H>10$ kOe they strongly segregate. Interestingly, the
characteristic value $H=10$ kOe is equal to the saturation field
$H_{\rm sat}^{\rm MN}$ of the MN. The behavior mentioned above is
observed only for the case where the field is normal to the film's
surface, while is absent when the field is parallel to the film.
Our experimental results suggest that the observed enhancement of
surface superconductivity field $H_{\rm c3}(T)$ is possibly due to
the not uniform local reduction of the external magnetic field by
the dipolar fields of the MN.

\end{abstract}

\pacs{74.25.Op, 74.78.Db, 74.25.Fy, 74.25.Dw}

\maketitle

\section{Introduction}

The interplay of superconductivity with magnetism is an
interesting subject of current theoretical and experimental
research. The interest on the experimental study of composite
superconductor-ferromagnet (SC/FM) structures is continuous not
only due to their importance for the further theoretical treatment
and understanding of the underlying mechanisms, but also because
of the promising applications that such combined structures could
give in the near future.\cite{Lange03,Helseth02} Such composite
structures are in the form of films consisting of SC/FM bilayers
or superlattices, \cite{Homma86,Bergeret03,Helseth02} periodic
nano or micro sized artificial FM dots or squares embedded in a
low-$T_c$ superconducting film,
\cite{Baert95,Martin99,Morgan98,Metlushko99,Martin00,Bael99,Stoll02,Terentiev00,Martin02}
etc.

In such artificial structures there are two basic mechanisms that
control the interaction between the superconducting order
parameter and the magnetic moments at the SC/FM interface. First
is the electromagnetic mechanism which is related to the
interaction of the superconducting pairs with the magnetic fields
induced by the FM component in the SC and second is the exchange
interaction that the superconducting pairs experience as they
enter the FM through the SC/FM interface.
\cite{Aladyshkin03,Buzdin03,Ginzburg56} This second mechanism
plays a crucial role when the SC and the FM layers are placed in
close proximity and the SC/FM interfaces are of high structural
quality. In such cases the so-called proximity effect dominates
and a number of interesting phenomena are revealed. In the
proximity effect the spin dependence of the exchange interaction
in the FM results in a favorable spin orientation for the
superconducting electrons. As a consequence the superconducting
pairs are destroyed as they are transmitted through the SC/FM
interface. Nevertheless, in many cases the superconducting pairs
may be reflected, rather than transmitted, at the SC/FM interface.
A phenomenological parameter called transparency is employed to
describe the transmission ability of the
interface.\cite{Aarts97,Aarts01} This parameter depends on the
structural quality of the interface and on the specific
spin-dependent scattering mechanism that the superconducting pairs
experience.\cite{Aarts97,Aarts01,Khusainov97,Lazar00,Baladie01,Cirillo}
In the case of very low transparency the superconducting pairs
never enter the FM as they are reflected at the SC/FM interface.
As a consequence the proximity effect is howsoever depressed. In
contrast, when nearly perfect interfaces are available their
transparency is very high and consequently the proximity effect is
fully restored with breaking of the transmitted pairs to occur in
the FM. Nevertheless, the whole process is not instant but occurs
at a time scale corresponding to a travelling length of the order
of the coherence length $\xi_F$ in the FM. Even in the clean limit
where $\xi_F=\hbar u_F/\Delta E_{ex}$ ($u_F$ is the Fermi velocity
and $\Delta E_{ex}$ is the exchange splitting) the characteristic
magnetic length is very small and as a result the superconducting
order parameter exhibits a rapid decay in the FM component. As a
consequence of its almost zeroing at the SC/FM interface the
superconducting order parameter should also be strongly depressed
at a scale of the order of the coherence length $\xi_S$ in the
SC.\cite{Gennes63,Demler97} Thus, it is generally expected that at
a SC/FM interface the proximity effect should be antagonistic to
the nucleation of superconductivity near the surfaces of the SC.

On the other hand when a SC is in proximity with an insulator (IN)
superconductivity should firstly appear not in the bulk of the SC
but at a thin layer of the order of $\xi_S$ near its
surfaces.\cite{James63} This is the so-called surface
superconductivity effect and occurs at a magnetic field $H_{\rm
c3}(T)$ which is higher than the bulk upper-critical field $H_{\rm
c2}(T)$ where superconducting order is established in the whole
sample.\cite{James63,Abrikosov88} As a consequence of the surface
superconducting layer for $H<H_{\rm c3}(T)$ the magnetoresistance
is lower than the normal state value and becomes zero only when
bulk superconductivity occurs i.e., at $H=H_{\rm c2}(T)<H_{\rm
c3}(T)$. In contrast, it is expected that in the regime $H_{\rm
c2}(T)<H<H_{\rm c3}(T)$ the measured magnetization of the
superconductor should be zero due to a change in the direction of
the superconducting screening currents that flow in the thin layer
near the surface of the sample.\cite{James63,Abrikosov88} Thus,
the effect of surface superconductivity can't be identified by
means of global magnetic measurements (SQUID measurements).

Until today, most of the reports on HS referred mainly to
transport properties performed in the regime just below the bulk
upper-critical field $H_{\rm c2}(T)$, or well inside the mixed
state of the superconductor and were limited in the low-field
regime close to the critical temperature. The influence of a
magnetic component to the transport behavior of the superconductor
in the high-field regime above the bulk upper-critical field
$H_{\rm c2}(T)$ and especially the behavior of the surface
superconductivity field $H_{\rm c3}(T)$ has not attracted the
interest of experimental studies although there are theoretical
reports related to the
subject.\cite{Cheng99,Buzdin03,Aladyshkin03} In this paper we
present detailed transport and magnetic data on the influence of
randomly distributed anisotropic CoPt MN on the superconducting
order parameter of the conventional low-$T_c$ Nb superconductor
which is a simple isotropic system, thus giving us the opportunity
to exclude parameters (thermal fluctuations or anisotropy) that
could complicate the total behavior of a HS. In the composite
system, we found that the surface superconductivity field $H_{\rm
c3}(T)$ is strongly enhanced in the high-field regime when the
applied field is normal to the film's surface. More specifically,
we found that although the related lines $H_{\rm c3}(T)$ of the HS
and pure Nb films coincide in the low-field regime $H<10$ kOe,
they strongly segregate for $H>10$ kOe. The characteristic value
$H=10$ kOe where this change occurs is equal to the saturation
field $H_{\rm sat}^{\rm MN}$ of the MN. In contrast, the bulk
upper-critical lines $H_{\rm c2}(T)$ of the pure Nb and the HS
films almost coincide. A comparison of our experimental results
with current theoretical knowledge is made and a simple
explanation is proposed for the observed behavior. We believe that
in our case the electromagnetic mechanism plays a dominant role,
while the proximity effect is probably depressed due to the low
transparency of the structurally disordered surfaces of the MN.
The high values that the dipolar field of the MN attains by their
lateral surfaces results in a sufficient reduction of the external
applied magnetic field in the respective regimes. As a result the
surface superconductivity field line $H_{\rm c3}(T)$ is strongly
enhanced in the HS film.

\section{Preparation of the films and Experimental details}

First of all, our aim was to produce pure Nb films. Relatively
clean films are needed in order to minimize the bulk pinning of
vortices. This is necessary in order to reveal the interaction
between the MN and the superconducting order parameter. The
sputtering conditions needed for the deposition of relatively
clean Nb films and for the preparation of the CoPt MN are reported
in Refs. \onlinecite{StamopoulosNb} and \onlinecite{Karanasos01}
respectively. The CoPt MN are isolated and randomly distributed on
the substrate's surface. Their magnetic behavior is anisotropic,
presenting a preferential out-of-plane magnetic moment. Their
typical in-plane size and thickness are of the order of $200$ nm
and $30$ nm respectively, while their distance is of the order of
$200$ nm (see transmission electron microscopy data in Ref.
\onlinecite{Karanasos01}). After producing the MN the Nb layer was
sputtered on top of them, so that the MN were actually embedded at
the bottom of the Nb layer (see fig. \ref{bnew} below). We should
underline that the pure Nb film and the Nb layer of the HS were
produced simultaneously during the same sputtering run since the
two substrates were mounted side by side. In this way the same
intrinsic properties of the Nb layer of the HS and of the pure Nb
film are obtained. The critical temperature of the films under
discussion is $T_c=8.3$ K for pure Nb and $T_c=7.7$ K for the HS.
The residual resistance ratio is RRR$\approx 3$. The determination
of the thickness of the Nb layer was based on an Alpha-Step device
and for the films under discussion was found to be $\approx 160$
nm ensuring that the SC is in the $3$D bulk limit. Typical in
plane dimensions of the films are $4\times 4$ mm$^2$. Combined
x-ray diffraction and transmission electron microscopy data
revealed that the crystallites of the Nb layer have mean size of
the order of $40$ nm and that they are homogeneous without
exhibiting columnar growth.

Our magnetoresistance measurements were performed by applying a dc
transport current (normal to the magnetic field) and measuring the
voltage in the standard four-point configuration. In most of the
measurements presented below the applied current was $I_{{\rm
dc}}=0.5$ mA, which corresponds to an effective density $J_{{\rm
dc}}\approx 80$ A/cm$^2$. The magnetic measurements were performed
under both zero field cooling and field cooling conditions. The
temperature control and the application of the dc fields were
achieved in a commercial SQUID device (Quantum Design). We
examined the whole temperature-magnetic-field regime accessible by
our SQUID ($H_{\rm dc}<55$ kOe, $T>1.8$ K).

\section{Experimental results and discussion}

\begin{figure}[tbp] \centering%
\includegraphics[angle=0,width=8cm]{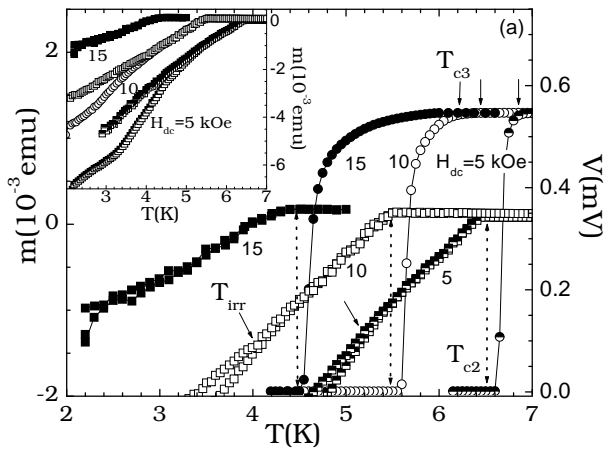}
\caption {The upper panel presents the comparison of the measured
voltage curves at a high current $I_{{\rm dc}}=0.5$ mA ($J_{{\rm
dc}}\approx 80$ A/cm$^2$) to magnetic data (circles and squares
respectively) for the HS as a function of temperature for magnetic
fields $H_{{\rm dc}}=5, 10$ and $15$ kOe. The voltage attains the
normal state value at a much higher temperature $T_{\rm c3}$ than
the upper-critical one $T_{\rm c2}$. The irreversibility points
$T_{\rm irr}$ (inclined arrows) are clearly distinct from the bulk
upper-critical temperatures $T_{\rm c2}$. In the inset we present
the magnetic measurements in an extended temperature regime. In
all cases the applied dc field is normal to the surface of the
film.}
\label{b0}%
\end{figure}

Figure \ref{b0}(a) presents a set of magnetic and resistance
measurements for magnetic fields $H_{{\rm dc}}=5, 10$ and $15$ kOe
normal to the surface of the HS film, while in the inset we
present the magnetic data in an extended temperature regime. In
the magnetic data we present both curves for zero field cooling
and subsequent field cooling. We observe that the bulk
upper-critical temperatures $T_{\rm c2}(H)$, defined from the
magnetic measurements as the point where the diamagnetic signal
becomes zero, coincide with the temperatures where for a high
applied current $I_{{\rm dc}}=0.5$ mA ($J_{{\rm dc}}\approx 80$
A/cm$^2$) the voltage starts taking nonzero values (see dotted
double arrows). In addition, the magnetically determined
irreversibility points $T_{\rm irr}(H)$ are clearly placed at much
lower temperatures comparing to $T_{\rm c2}(H)$. This fact
indicates that the identified upper-critical points $T_{\rm
c2}(H)$ refer to the zeroing of the equilibrium magnetization and
not to a collapse of screening currents that originate from bulk
pinning. More importantly, we clearly observe that the voltage
curves take the normal state value at temperatures much higher
than $T_{\rm c2}(H)$. Furthermore, the $I-V$ characteristics
exhibited a clear non-linear behavior in the regime above $T_{\rm
c2}(H)$ (see below). According to currently available knowledge
the only characteristic effect placed outside the mixed state of a
type-II superconductor ($T>T_{\rm c2}(H)$) is surface
superconductivity. Thus, we identify the points where the
resistance attains the normal state value as the characteristic
temperatures $T_{\rm c3}(H)$ where surface superconductivity
occurs.

\begin{figure}[tbp] \centering%
\includegraphics[angle=0,width=8cm]{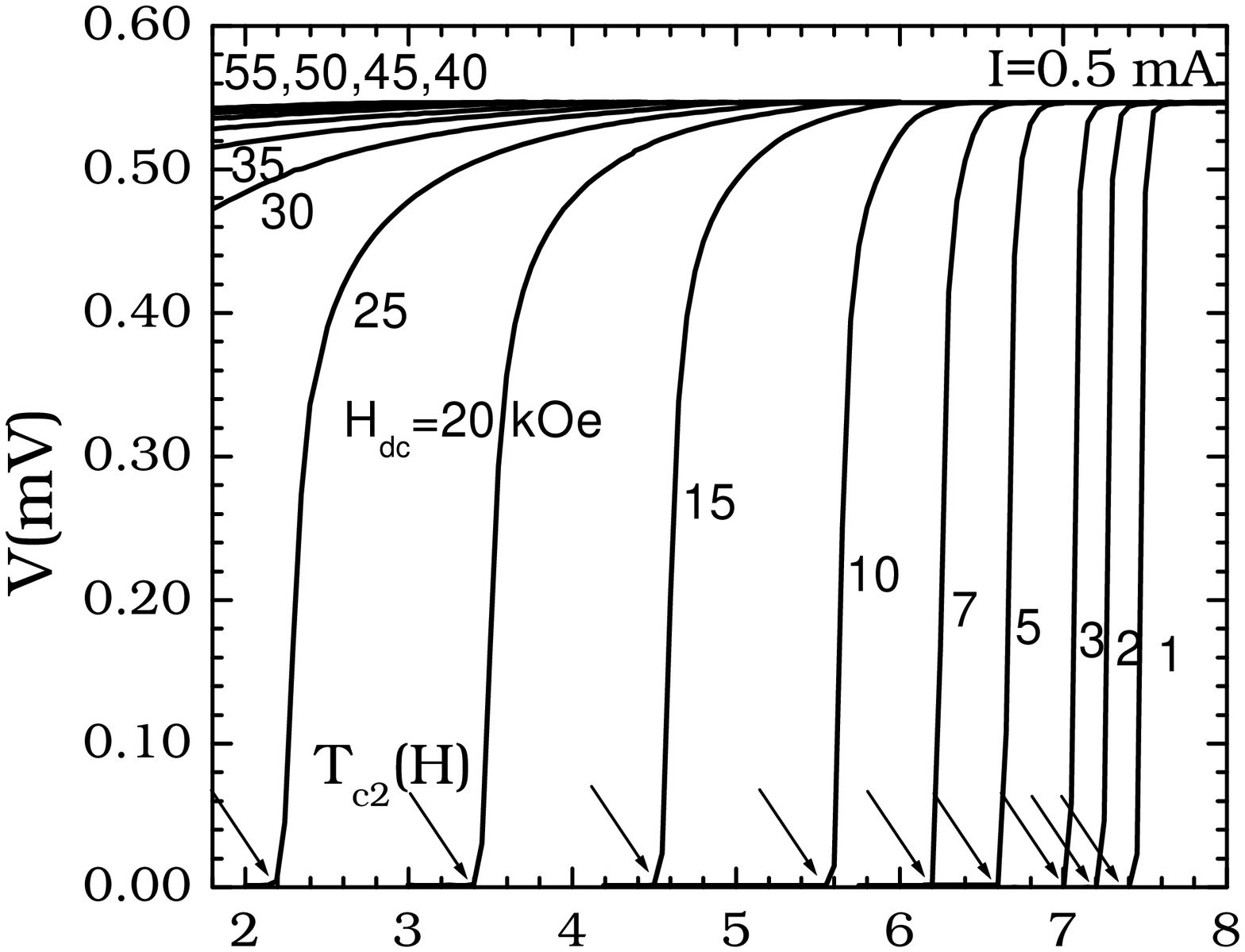}
\includegraphics[angle=0,width=8cm]{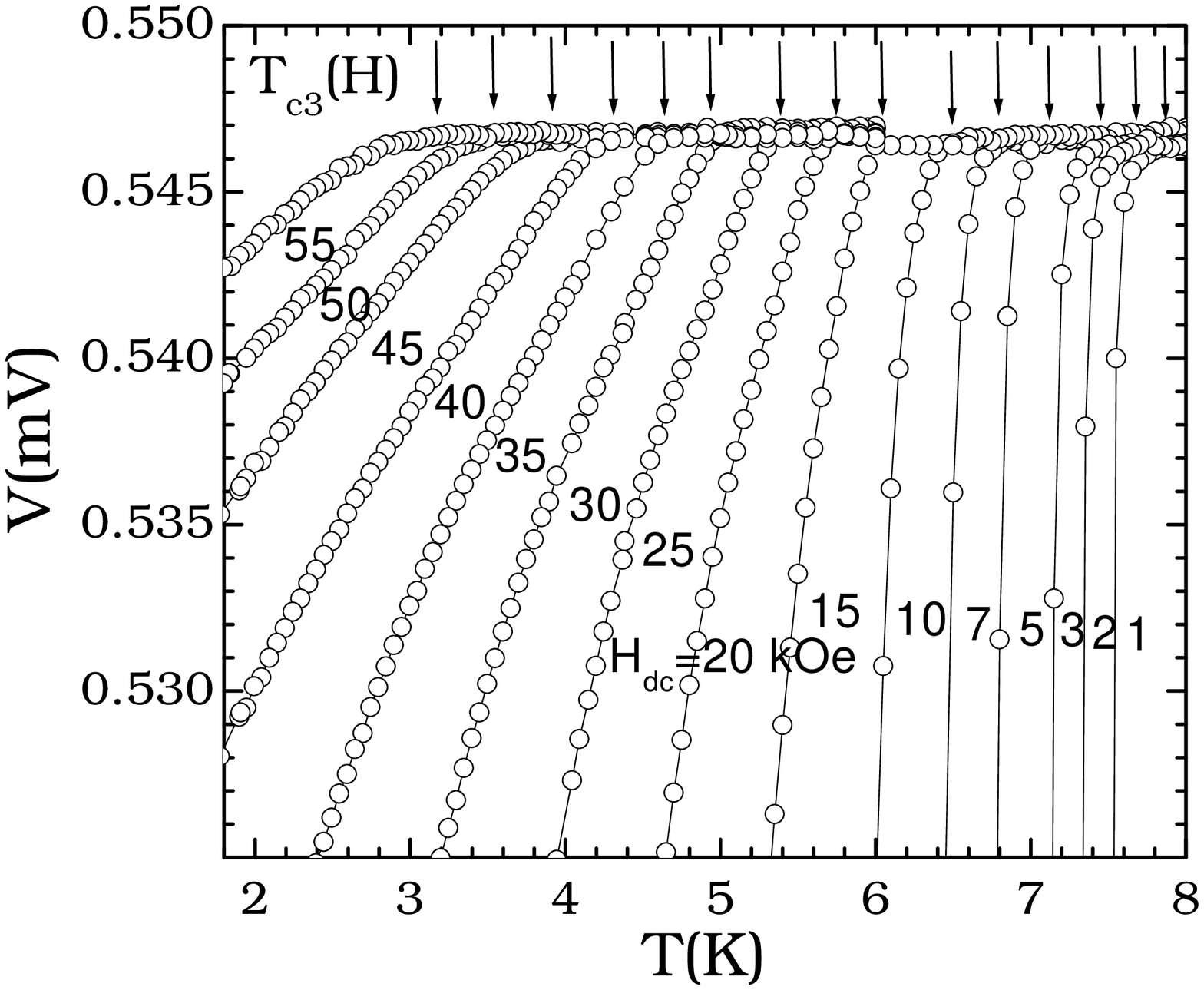}
\caption { Measured voltage in the HS as a function of temperature
for various dc magnetic fields under a dc transport current
$I_{{\rm dc}}=0.5$ mA ($J_{{\rm dc}}\approx 80$ A/cm$^2$) when the
dc field is normal to the surface of the film. The upper panel
presents the whole resistive transition, while in the lower panel
we focus in the regime where the voltage attains its normal state
value. }
\label{b1}%
\end{figure}%

\begin{figure}[tbp] \centering%
\includegraphics[angle=0,width=8cm]{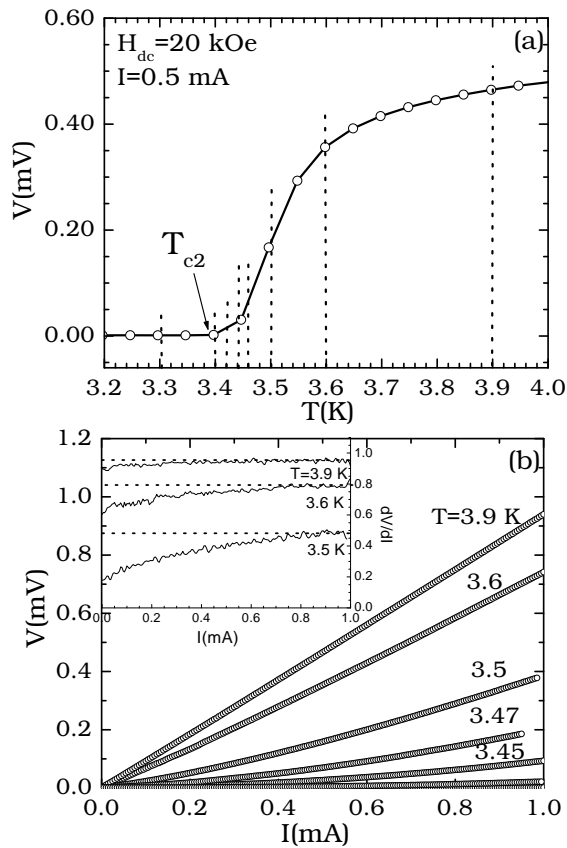}
\caption { Measured voltage in the HS as a function of temperature
for $I_{{\rm dc}}=0.5$ mA (upper panel) and representative $I-V$
characteristics (lower panel) at the temperatures indicated by the
dotted lines. The data refer to $H_{{\rm dc}}=20$ kOe normal to
the surface of the film. The inset presents the derivative of the
measured $I-V$ curves for $T=3.5, 3.6$ and $3.9$ K. Even at
$T=3.9$ K the $I-V$ curve is non-linear in low applied currents.}
\label{bb1}%
\end{figure}

In figs. \ref{b1}(a) and \ref{b1}(b) we present a complete set of
magnetoresistance measurements in the HS film at a constant
current $I_{{\rm dc}}=0.5$ mA ($J_{{\rm dc}}\approx 80$ A/cm$^2$)
for various magnetic field values. We observe that the voltage
curves are strongly rounded. The observed rounding gradually
disappears as we apply lower magnetic fields. In addition, we see
that the points $T_{\rm c3}(H)$ where the voltage takes the normal
state value are well resolved in our measurements (see fig.
\ref{b1}(b)) and consequently may be defined with high accuracy.
In order to investigate the regime $T_{\rm c2}(H)<T<T_{\rm c3}(H)$
in more detail we also performed measurements as a function of the
applied current. Representative $I-V$ characteristics are shown in
fig. \ref{bb1} for various temperatures at a magnetic field
$H_{{\rm dc}}=20$ kOe. In the upper panel the dotted lines
indicate the specific temperatures where the measurements have
been performed, while the lower panel presents the respective
data. For $T<T_{\rm c2}(H)=3.4$ K the maximum applied current
$I_{{\rm dc}}=1$ mA ($J_{{\rm dc}}\approx 160$ A/cm$^2$) doesn't
exceed the bulk critical current and the detected voltage is
almost zero. Above $T_{\rm c2}(H)=3.4$ K the applied current
exceeds the surface critical current and a finite voltage is
measured. The observed $I-V$ curves are non-linear in the regime
$T_{\rm c2}(H)<T<T_{\rm c3}(H)$ and become linear only when
$T_{\rm c3}(H)$ is exceeded. In the inset we present the
derivative of the measured $I-V$ curves. We see that even at
$T=3.9$ K the response is non-linear in small applied currents. In
agreement to other works the experimental data presented in fig.
\ref{bb1} give evidence that the regime $T_{\rm c2}(H)<T<T_{\rm
c3}(H)$ is governed by surface superconductivity.
\cite{Hempstead64,DeSorbo64B,James63,Abrikosov88,StamopoulosNb,Welp03,Rydh03}
Although the original theory predicted the presence of the effect
only for the case where the field is parallel to the surface of
the specimen, recent experimental reports investigated also the
case where the field is normal to the surface of the sample giving
evidence that an analogous effect takes place even for such
configuration. \cite{Welp03,StamopoulosNb,Rydh03}

\begin{figure}[tbp] \centering%
\includegraphics[angle=0,width=8cm]{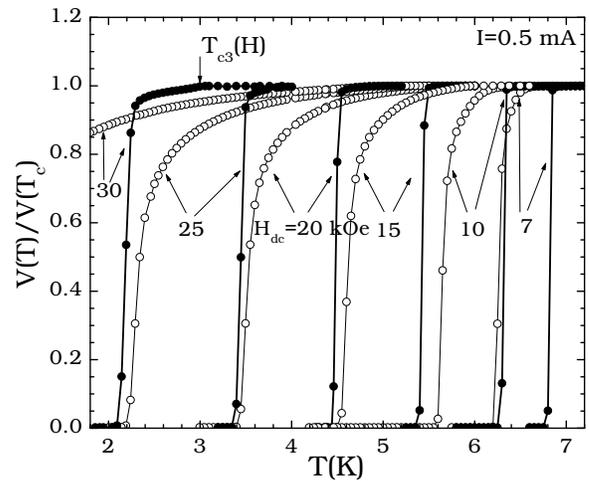}
\caption { Measured voltage on the HS film as a function of
temperature for various dc magnetic fields under a dc transport
current $I_{{\rm dc}}=0.5$ mA ($J_{{\rm dc}}\approx 80$ A/cm$^2$)
when the dc field is normal (open circles) and when is parallel
(solid circles) to the surface of the film. In both cases the
transport current is transverse to the magnetic field. }
\label{b2}%
\end{figure}

Since the MN exhibit anisotropic magnetization, different
behaviors should be observed when the magnetic field is normal or
parallel to the HS film. In fig. \ref{b2} we comparatively present
such measurements for both field configurations. At the first set
the field was normal (open circles) while at the second set was
parallel (solid circles) to the surface of the film. In both sets
the dc transport current was transverse to the magnetic field. As
we clearly see the resistive transition for the case where the
field is normal to the surface of the HS film is rounded and more
extended compared to the one for the case where the field is
parallel to the film's surface.

\begin{figure}[tbp] \centering%
\includegraphics[angle=0,width=8cm]{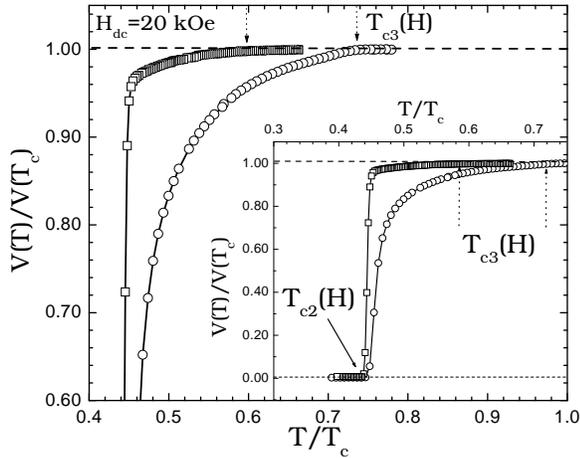}
\caption {Comparison of the measured normalized voltage for the
pure Nb (squares) and the HS (circles) films as a function of
reduced temperature under a dc transport current $I_{{\rm
dc}}=0.5$ mA ($J_{{\rm dc}}\approx 80$ A/cm$^2$) for $H_{{\rm
dc}}=20$ kOe. In both measurements the magnetic field is normal to
the surface of the film. }
\label{b3}%
\end{figure}

\begin{figure}[tbp] \centering%
\includegraphics[angle=0,width=8cm]{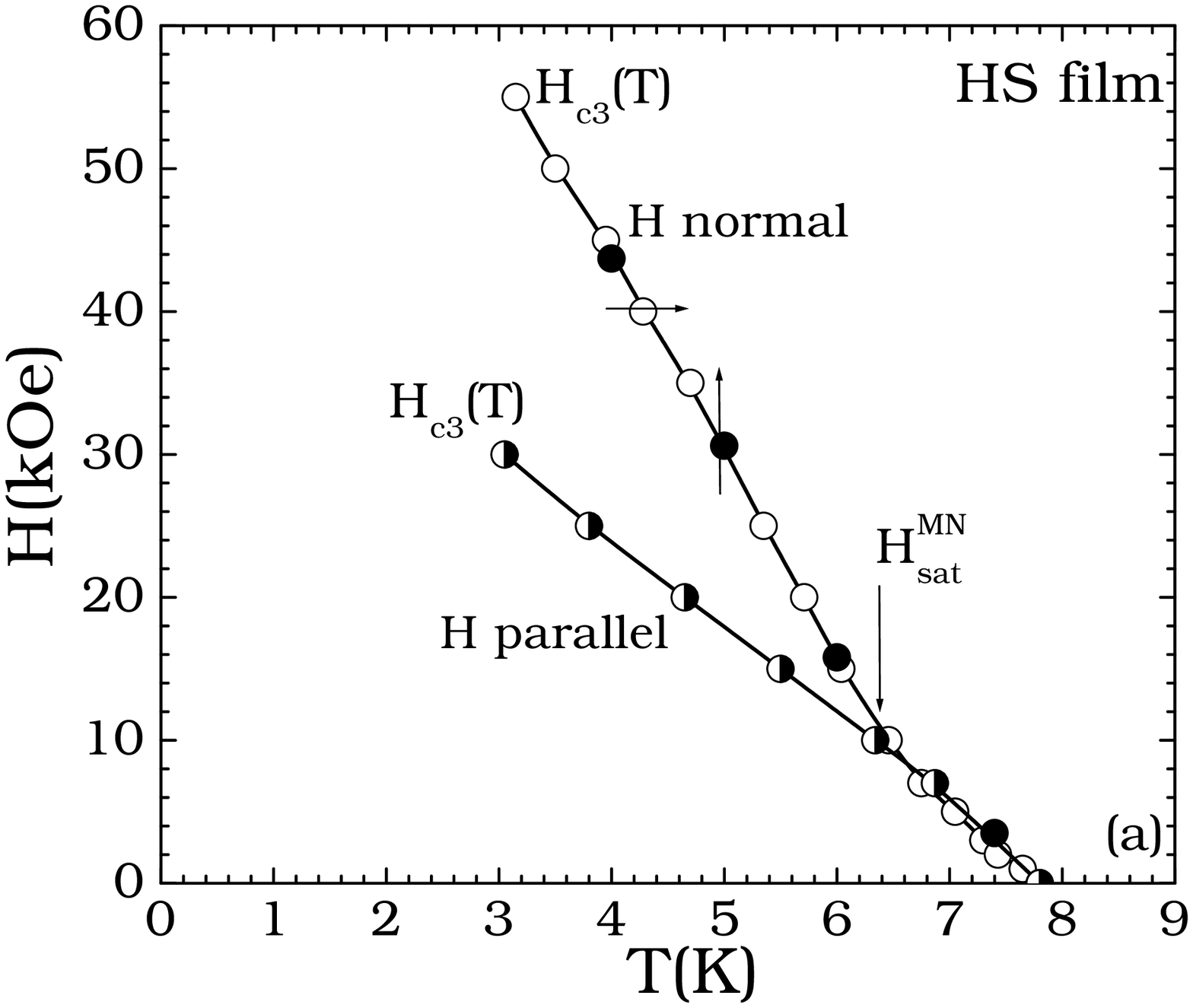}
\includegraphics[angle=0,width=8cm]{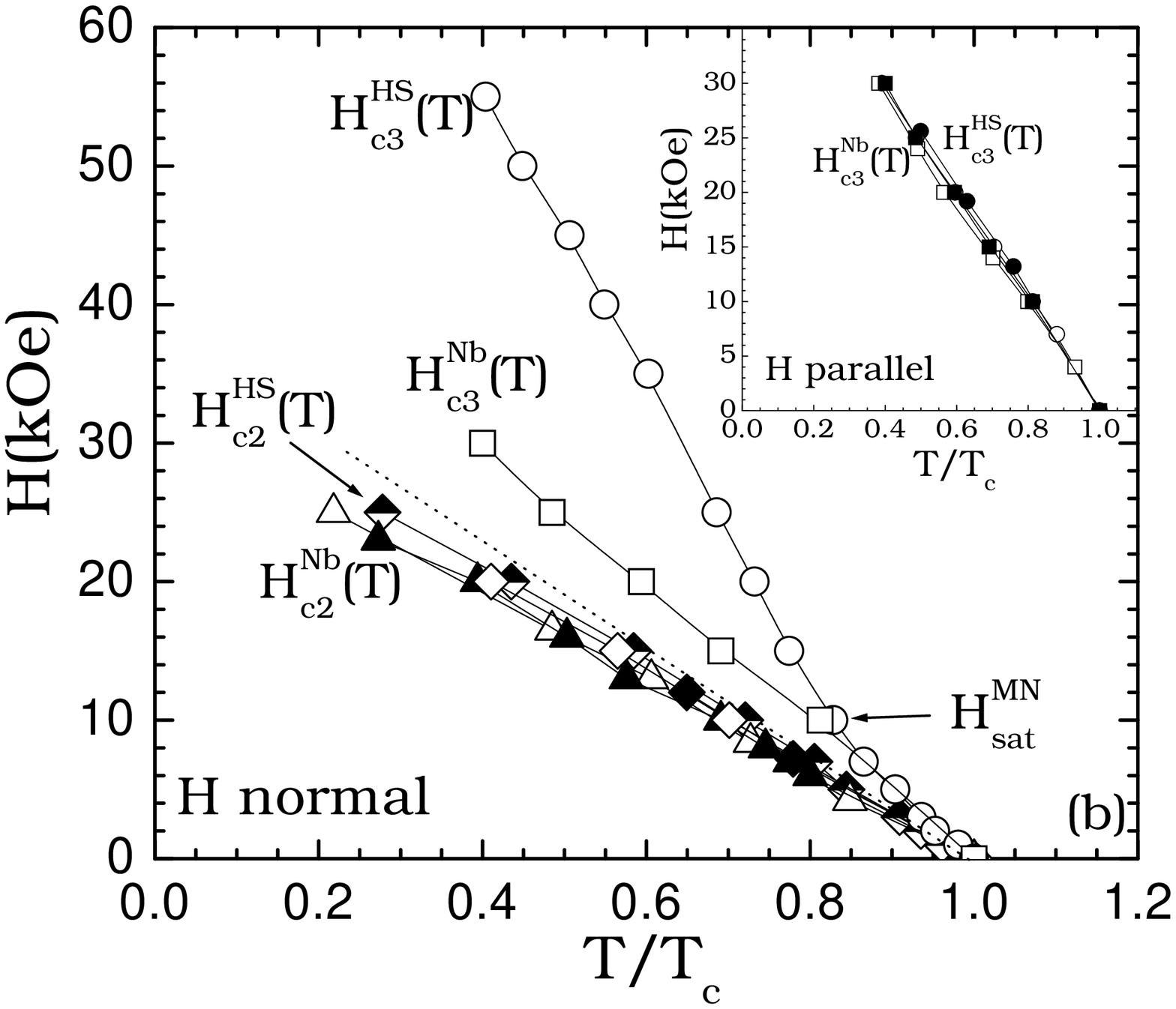}
\caption { The upper panel presents the characteristic field lines
$H_{\rm c3}(T)$ where the voltage attains its normal state value
for the HS film measured when the magnetic field is normal (open
and solid circles) and when is parallel (semi-filled circles) to
the surface of the film. In the lower panel and in reduced
temperature presented are the upper-critical field lines $H_{\rm
c2}(T)$ (triangles and rhombi for the Nb and the HS films
respectively), and the characteristic fields $H_{\rm c3}(T)$ where
the voltage attains its normal state value, of the pure Nb
(squares) and HS (circles) films measured when the magnetic field
is normal to the surface of the film. In the inset we present the
$H_{\rm c3}(T)$ lines for the pure Nb (squares) and for the HS
(circles) films measured when the magnetic field is parallel to
the surface of the film. In all cases the transport current is
transverse to the magnetic field. }
\label{b4}%
\end{figure}

In fig. \ref{b3} we present in reduced temperature units a
comparison of the voltage measured in pure Nb and HS films under a
magnetic field $H_{\rm dc}=20$ kOe normal to their surfaces. We
see that the upper-critical points $T_{\rm c2}$($20$ kOe) almost
coincide (see the inset) but the voltage attains its normal state
value at a much higher temperature $T_{\rm c3}$($20$ kOe) for the
HS film compared to the pure Nb one. On the other hand, the
corresponding measurements performed for the configuration where
the applied magnetic field is parallel to the main surface of the
films exhibited a totally different behavior. In the parallel
configuration the $T_{\rm c3}$($20$ kOe) points of the HS and of
the pure Nb films coincide when plotted in reduced temperature
units (see the inset of fig. \ref{b4}(b) below).

The experimental results presented above are summarized in figs.
\ref{b4}(a) and \ref{b4}(b). Figure \ref{b4}(a) presents the
surface superconductivity fields $H_{\rm c3}(T)$ for the HS film
for the case where the field is normal (open and solid circles)
and parallel (semi-filled circles) to the surface of the film. The
open (solid) circles come from isofield (isothermal) measurements
as a function of temperature (field). We see that for the two
field configurations the characteristic lines $H_{\rm c3}(T)$
clearly coincide up to a certain magnetic field $H\approx 10$ kOe,
but above this value the two lines strongly diverge. Figure
\ref{b4}(b) summarizes the main result of the present study
referring to the configuration where the {\it field is normal} to
the film's surface. The $H_{\rm c2}(T)$ and the $H_{\rm c3}(T)$
lines of the HS and of the pure Nb films are presented in reduced
temperature units. The $H_{\rm c2}(T)$ lines are determined by
magnetic (open and solid rhombi for the HS film) and
magnetoresistance measurements (triangles and semi-filled rhombi
for the pure Nb and HS films respectively). The most important
result presented in fig. \ref{b4}(b) is that the characteristic
lines $H_{\rm c3}(T)$ coincide up to $H\approx 10$ kOe but above
this field they strongly segregate, with the $H_{\rm c3}(T)$ line
of the HS film placed at much higher fields comparing to the
respective one of pure Nb. The characteristic value $H\approx 10$
kOe is equal to the saturation field of the MN. This experimental
fact gives strong evidence that the MN are responsible for the
behavior observed in $H_{\rm c3}(T)$ (see below). Furthermore,
above $H\approx 10$ kOe the bulk upper-critical field line $H_{\rm
c2}(T)$ of the HS presents a smooth deviation from the
extrapolated high temperature data (see dotted line). This
behavior is not observed in the $H_{\rm c2}(T)$ data of pure Nb.
Thus, regarding $H_{\rm c2}(T)$ the preliminary results presented
in this work suggest that the weak rounding observed in the line
$H_{\rm c2}(T)$ of the HS is motivated by the MN (as is the
pronounced one observed in the line $H_{\rm c3}(T)$). Despite
that, this gradual deviation could be also ascribed to the common
zeroing of its slope as the line $H_{\rm c2}(T)$ of a disordered
superconductor approaches zero temperature. In this work we focus
on the pronounced effect observed in $H_{\rm c3}(T)$ and we won't
refer further on the weak rounding observed in $H_{\rm c2}(T)$
since more experiments are needed to clarify this point. By
extrapolating the related data to zero temperature we find that
the ratio of the surface superconductivity field $H_{\rm c3}(T)$
to the bulk upper-critical field $H_{\rm c2}(T)$ is $H_{\rm
c3}(0)/H_{\rm c2}(0)$=$1.5$ and $2.9$ for the pure Nb and the HS
films respectively. The estimated value $H_{\rm c3}(0)/H_{\rm
c2}(0)$=$1.5$ for the pure Nb film is close to the theoretically
proposed value of $1.695$.\cite{James63,Abrikosov88} In contrast,
the corresponding value $H_{\rm c3}(0)/H_{\rm c2}(0)$=$2.9$ for
the HS film is strongly increased. For the case where the magnetic
{\it field is parallel} to the surface of the films the situation
is entirely different as presented in the inset in reduced
temperature units. We observe that the lines $H_{\rm c3}(T)$ of
the pure Nb (squares) and HS (circles) films almost coincide in
the whole temperature-magnetic-field regime investigated in this
work.

As we mentioned in the introduction there are two basic mechanisms
that control the interaction between the superconducting order
parameter and the magnetic moments at a SC/FM bilayer. First is
the electromagnetic mechanism and second is the exchange
interaction that the superconducting pairs experience as they
enter the FM through the SC/FM interface. The second mechanism
results in a strong depression of the superconducting order
parameter at the surfaces of a SC which are placed in close
proximity to the FM component (proximity
effect).\cite{Aladyshkin03,Buzdin03,Ginzburg56} In contrast, at a
SC/IN interface the surface superconductivity effect implies that
for magnetic fields $H_{\rm c2}(T)<H<H_{\rm c3}(T)$ the
superconducting order parameter is enhanced near the surfaces of
the SC. Consequently, the so-called proximity effect should act
against the formation of surface superconductivity. Our data give
evidence for an enhancement of surface superconductivity in the HS
film. Thus, we speculate that probably the proximity effect
doesn't have strong influence in our case. One way to eliminate
the proximity effect is by placing a sufficiently thick insulating
spacer between the SC and the FM components. In such structures
the insulating layer ensures that the superconducting pairs don't
experience the dominant exchange interactions as they never enter
the FM. On the other hand, recent studies
\cite{Aarts97,Aarts01,Khusainov97,Lazar00,Baladie01,Cirillo}
showed that an important phenomenological parameter called
transparency should be introduced in order the theory to be
consistent with the experimental results. This parameter which
describes the transmission ability of the SC/FM interface depends
on extrinsic and intrinsic factors. The main extrinsic
contribution comes from the structural quality of the SC/FM
interface (compositional disorder, oxidation, mismatch of the
lattices that results in strain effects),
\cite{Aarts97,Lazar00,Baladie01,Cirillo} while the intrinsic one
is related to the specific spin dependent scattering mechanism
that the superconducting pairs experience as go through the
interface.\cite{Aarts97,Lazar00,Jong95} For the case of perfect
SC/FM surfaces the transparency is maximum and as a result the
proximity effect may be easily observed. In contrast, strongly
disordered interfaces are described by low transparency and the
proximity effect could be strongly depressed. Due to the
preparation technique employed in our work the produced MN don't
have flat surfaces, in contrast to other works where the magnetic
dots or squares were usually fabricated by means of a controllable
lithographic technique. Thus, in our case the proximity effect
could be strongly depressed due to structural distortion and
possible oxidation of the SC/FM interfaces.

In order to leave no doubt, below we compare our experimental
results with recent theoretical studies \cite{Takahashi86A} and
experimental works
\cite{Kanoda86,Banerjee83,Chun84,Sidorenko96,Hettinger96,Homma86,Jiang96}
that refer to similar phenomena motivated by the proximity effect
in superconducting-normal metal (SC/NM) or SC/FM multilayers. In
recent studies Takahashi and Tachiki \cite{Takahashi86A} suggested
that due to the proximity effect a strong deviation in the bulk
upper-critical field $H_{\rm c2}(T)$ should be observed in a SC/NM
multilayer. A brief summary of their conclusions is given here.
The constituent layers of the superlattice are characterized by
their diffusion constants $D$, pairing potentials $V$, electronic
density of states $\Lambda$ and thicknesses $d$. By varying each
parameter separately the authors \cite{Takahashi86A} showed that
both parallel and normal bulk upper-critical fields exhibit an
intense change at some characteristic temperature. First, let us
suppose that only $\Lambda$ is discontinuous at the SC/NM
interface, while $D_S=D_N$, $V_S=V_N$ and $d_S=d_N$. For
temperatures close to $T_c$ the usual three dimensional linear
behavior is observed in the parallel bulk upper-critical field
$H_{\rm \|c2}(T)\propto(1-T/T_o)$. When decreasing the temperature
the superconducting coherence length $\xi_S$ also decreases and at
a characteristic temperature it becomes equal both to $d_S$ and
$d_N$. Below this characteristic temperature, $H_{\rm \|c2}(T)$
exhibits a square-root dependence
$H_{\|c2}(T)\propto(1-T/T_o)^{1/2}$ which is indicative of a two
dimensional behavior. Thus, in the low temperature regime the
layers of the SC become decoupled. The same qualitative behavior
is also observed in $H_{\rm \|c2}(T)$ when the conditions
$\Lambda_S=\Lambda_N$, $V_S=V_N$ and $d_S=d_N$ hold, but the
diffusion constants of the constituent layers are unequal $D_S\neq
D_N$. In the later case the effect is even more dramatic for the
normal upper-critical field $H_{\rm \perp c2}(T)$. While close to
the critical temperature the $H_{\rm \perp c2}(T)$ exhibits the
usual linear behavior, below a characteristic temperature acquires
a pronounced upturn. It is worth noting that in contrast to
$H_{\rm \| c2}(T)$ the $H_{\rm \perp c2}(T)$ exhibits almost
linear behavior in a wide regime both below and above the
characteristic temperature. This is reminiscent of the behavior
observed in our data. Thus, a comparison of the two cases is
necessary. Due to the fact that the CoPt MN are embedded at the
bottom of the Nb film a normal applied magnetic field could be
experienced as parallel by the sequential (but not periodic)
lateral MN/Nb interfaces (see fig. \ref{bnew} below). So, the
present experimental configuration could be analogous, in some
degree, to the theoretically described \cite{Takahashi86A} and
experimentally studied artificial SC/NM
\cite{Kanoda86,Banerjee83,Chun84,Sidorenko96,Hettinger96} or SC/FM
\cite{Homma86,Jiang96} multilayers. Interestingly, the shape of
the resulting line $H_{\rm c3}(T)$ observed in our case is similar
to the one exhibited by the {\it parallel} ''upper-critical
field'' $H_{\rm \| c2}(T)$ when a crossover from three dimensional
to two dimensional behavior was observed in the above mentioned
experimental and theoretical reports.
\cite{Kanoda86,Takahashi86A,Banerjee83,Chun84,Sidorenko96,Hettinger96,Homma86,Jiang96}
Despite this similarity there are also strong differences that
should be mentioned. First of all, we have to underline that in
those works absolutely {\it periodic} multilayered structures were
employed, while in our case we have a non periodic distribution in
the MN/Nb interfaces. Second, our $H_{\rm c3}(T)$ data don't
exhibit a $(1-T/T_o)^{1/2}$ dependence in the low temperature
regime. Third and more important, our experimental data refer to
the surface superconductivity field line $H_{\rm c3}(T)$ and not
to the bulk upper-critical fields $H_{\rm c2}(T)$. Due to the
above arguments we believe that either the behavior observed in
our $H_{\rm c3}(T)$ data is not influenced by the proximity
effect, or the results of previously published experimental works
refer also to the surface superconductivity field $H_{\rm c3}(T)$.
The second case is possible since in most works the upper-critical
field was usually defined by the midpoints of the resistive
transition without comparing with magnetic measurements. As a
consequence, the data presented in the past could more or less
detect $H_{\rm c3}(T)$ and not $H_{\rm c2}(T)$. Finally, we note
that until today there is no theoretical work investigating the
problem of a thick SC layer in proximity with a FM in the field
range $H_{\rm c2}(T)<H<H_{\rm c3}(T)$ where surface
superconductivity is expected in the SC. The antagonistic role of
the proximity effect and surface superconductivity strongly
complicates this problem. Our experimental results could give some
information for future theoretical treatment of this problem.

\begin{figure}[tbp] \centering%
\includegraphics[angle=0,width=8.5cm]{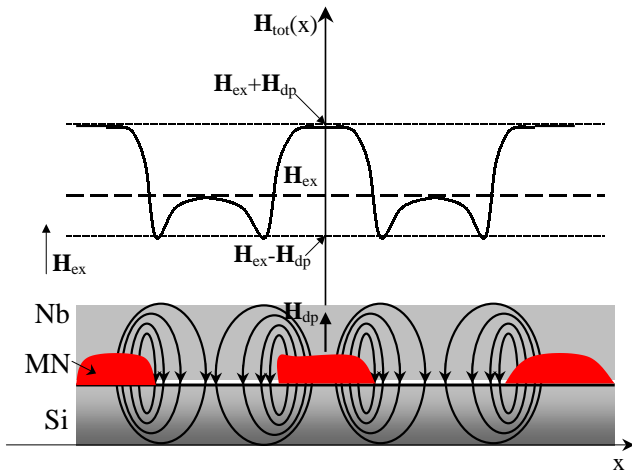}
\caption {Schematic presentation of the HS consisting of the MN
embedded at the bottom of the Nb layer. The external applied field
${\bf H}_{\rm ex}$ (thick dashed line) is modulated by the dipolar
fields ${\bf H}_{\rm dp}$ of the MN. Due to the inhomogeneous
intensity of the dipolar fields, the total local field ${\bf
H}_{\rm tot}$ (thick solid curve) presents a local minimum very
close to the lateral surfaces of the MN. Away from the MN the
total local field equals the external field.}
\label{bnew}%
\end{figure}%

By assuming that in our case the electromagnetic mechanism
dominates over the proximity effect, in the discussion given below
we propose a simple explanation for our experimental results.
Magnetization loop measurements performed in the HS film in
temperature just above $T_c$ (not shown here) revealed that the
saturation field of the MN is $H_{\rm sat}^{\rm MN}\approx 10$ kOe
(see also Ref. \onlinecite{Karanasos01}). This value is equal to
the characteristic field above (below) which the respective lines
$H_{\rm c3}(T)$ of the HS and the pure Nb films diverge (coincide)
when the field is normal to the film's surface. We believe that
this {\it purely experimental fact} may lead us to the following
qualitative interpretation for the behavior observed in figs.
\ref{b4}(a) and \ref{b4}(b). Under zero magnetic field the
anisotropic MN have their moments randomly distributed so their
macroscopic magnetic moment is zero. When we apply the magnetic
field normal to the film we also align the dipolar fields of the
MN normal to the film's surface. The situation discussed here is
schematically presented in fig. \ref{bnew}. The dipolar fields
${\bf H}_{\rm dp}$ will reduce (enhance) the external applied
field ${\bf H}_{\rm ex}$ in the region by the lateral surfaces of
the MN (in the region above their normal surfaces), so that the
total effective magnetic field ${\bf H}_{\rm tot}$ is lower
(higher) than the external magnetic field in the respective
regimes. As our experimental results showed, the whole effect
takes place clearly above the bulk upper-critical-field line
$H_{\rm c2}(T)$, so that the dipolar fields ${\bf H}_{\rm dp}$ of
the MN are not efficient to reduce the external applied field
${\bf H}_{\rm ex}$ in the bulk of the superconductor and drive it
in the mixed state. More importantly, we believe that the observed
effect is related to the fine local {\it uneven} distribution in
the intensities of the dipolar fields ${\bf H}_{\rm dp}$ (see fig.
\ref{bnew}). The dipolar fields of the MN are not uniform in the
surrounding space but have higher intensity in the vicinity close
to their lateral surfaces, while away from them decay rapidly to
zero. This could lead to a more efficient compensation of the
external magnetic field in the vicinity by the lateral surfaces of
the MN. Such a reduction of the external field could enhance the
formation of a superconducting condensate confined close to the
lateral surfaces of the MN, thus promoting surface
superconductivity.

Finally, we discuss another possible mechanism that could motivate
the rounding observed in our resistance curves. Due to their
preparation procedure the MN are not identical but a distribution
in their sizes exist. Although the distribution of their in-plane
dimensions could possibly assist a rounding in the measured
voltage curves, it couldn't be the main underlying mechanism that
motivates the observed effect. A distribution in their
out-of-plane dimension (height) also exists and this should also
lead to the same effect when the field is parallel to the film's
surface. This is not observed in our results. Consequently, we
believe that the distribution in the dimensions of the MN couldn't
be the main reason for the observed effect.

\section{conclusions}

In conclusion, we presented detailed transport and magnetic
measurements in pure Nb and in HS films in the whole
temperature-magnetic-field regime accessible by our SQUID. We
observed that when the magnetic field is parallel to their
surfaces the characteristic lines $H_{\rm c3}(T)$ coincide
entirely when plotted in reduced temperature units. In contrast,
the situation changes strongly when the field is normal to the
surface of the films. Although that for this field configuration
the bulk upper-critical fields $H_{\rm c2}(T)$ of the two films
almost coincide in the entire temperature regime, their respective
characteristic lines $H_{\rm c3}(T)$ coincide up to the saturation
field $H_{\rm sat}^{\rm MN} \approx 10$ kOe of the MN, while above
this field they strongly segregate. In our case the proximity
effect is possibly depressed due to the strongly distorted
surfaces of the MN. More importantly, recent works suggest that
the proximity effect has a strong influence on the bulk
upper-critical fields $H_{\rm c2}(T)$, while our experimental
results refer to the surface superconductivity field $H_{\rm
c3}(T)$. Thus we believe that in our case the electromagnetic
mechanism is dominant and motivates the observed behavior.
Consequently, we propose that in our HS film the enhancement of
surface superconductivity could be ascribed to the reduction of
the external applied field by the uneven dipolar fields of the MN
in the vicinity by their lateral surfaces. We hope that our
experimental results will promote future theoretical studies on
the possible coexistence of surface superconductivity and exchange
interactions at a bilayer of a thick SC placed in proximity with a
FM.


\pagebreak

\end{document}